\documentclass[twocolumn,twoside]{IEEEtran}
\usepackage{amsfonts}

\begin{document}

\title{A Public-key based Information\\Management Model for Mobile Agents}

\author{Diego Rodr\'{\i}guez and Igor Sobrado\thanks{ D. Rodr\'{\i}guez is
with the Technical School of Computer Sciences (EUITIO), University of
Oviedo, Oviedo (Asturias), Spain. E-mail: diego@string1.ciencias.uniovi.es .
I. Sobrado is with the Computer Sciences Department, University of
Oviedo, Oviedo (Asturias), Spain. E-mail: sobrado@acm.org . Authors' names
appear in alphabetical order.}\thanks{This work has been submitted to the
IEEE for possible publication. Copyright may be transferred without
notice, after which this version may no longer be accessible.}}

\markboth{IEEE Transactions On Networking, Vol. XX, No. Y, Month 2000}
{Rodr\'{\i}guez and Sobrado: A Public-key based Information Management
Model for Mobile Agents}

\maketitle

\begin{abstract}
Mobile code based computing requires development of protection schemes
that allow digital signature and encryption of data collected by the
agents in untrusted hosts. These algorithms could not rely on carrying
encryption keys if these keys could be stolen or used to counterfeit data
by hostile hosts and agents. As a consequence, both information and keys
must be protected in a way that only authorized hosts, that is the host
that provides information and the server that has sent the mobile agent,
could modify (by changing or removing) retrieved data. The data management
model proposed in this work allows the information collected by the agents
to be protected against handling by other hosts in the information
network. It has been done by using standard public-key cryptography
modified to support protection of data in distributed environments without
requiring an interactive protocol with the host that has dropped the
agent. Their significance stands on the fact that it is the first model
that supports a full-featured protection of mobile agents allowing remote
hosts to change its own information if required before agent returns to
its originating server.
\end{abstract}


\section{Introduction}

\PARstart{G}{lobal-networking} and world-wide communications are changing
computing concepts as were established some years ago. In the past,
information was stored in trusted hosts where data was managed. In that
case, it is easy to see that information security depends on the
protection of the hosts themselves. Protection of hosts should be
considered as a part of the operating systems design and not as an add-on.
In fact, to ignore last approach opens a wide variety of security holes in
modern computer systems \cite{ritchie:security}. In any case, hosts
protection is more advanced that mobile code protection at present time.
Distributed computing requires information to be managed in untrusted
---and sometimes unknown--- hosts in the network; consequently, new data
management models must be developed for distributed environments
\cite{hohl:mess,sander:hosts,sobrado:otp}. Some protection schemes based
in the use of Partial Result Authentication Codes (PRACs) ensures a
\textit{perfect forward integrity\/} but does not assures backward
integrity \cite{yee:sanctuary}. In these cases, the mobile agent carries
the keys that will be used to protect the messages. Each key could be
applied to protect only one message being removed from the agent data area
after using. Even forward integrity could be compromised. For example
suppose that an agent follows a route ${\cal H} = \{{\rm H}_{1},
{\rm H}_{2}, \ldots, {\rm H}_{i}^{\rm \ast}, {\rm H}_{i+1}, \ldots,
{\rm H}_{j-1}, {\rm H}_{j}^{\rm \ast}, \ldots, {\rm H}_{n}\}$, where both
${\rm H}_{i}^{\rm \ast}$ and ${\rm H}_{j}^{\rm \ast}$, with $i, j \in \{
1, 2, \ldots, n \}$ and $i < j$, are malicious hosts. The first hostile
node in the route of the agent, that is ${\rm H}_{i}^{\rm \ast}$, could
provide a copy of the keys not removed from the mobile agent to the second
malicious host, ${\rm H}_{j}^{\rm \ast}$. The second host could use this
information to counterfeit data provided by the hosts that apply these
keys. This fact makes all the hosts in the subset ${\cal H'} = \{
{\rm H}_{i+1}, {\rm H}_{i+2}, \ldots, {\rm H}_{j-1}\} \subset {\cal H}$
vulnerable. The same problem will happen if the agent returns to the first
malicious host later, showing that backward integrity cannot be assured in
any case.

The term \textit{message\/}, as applied in this paper, stands for each
block of information provided by the remote hosts to either other hosts or
mobile agents. The word \textit{field\/} identifies a chunk of the message
itself.

We want to note some differences that can be found between mobile agents,
mobile code and intelligent agents, to show in detail the problem that
could be solved by using the threat described in our work:

\begin{itemize}
\item\textit{Mobile agents.\/} A mobile agent is a software object that
have a code area and a data part. Both code and data areas will convey
from a host to another one but the execution thread will not be preserved.
Mobile agents could be easily implemented by using serialization in
programming languages as Java.

\item\textit{Mobile code.\/} The most important difference between mobile
agents and mobile code is that the latter allows the execution thread to
be preserved when the agent goes from one host to the next one. As the
execution thread will be changed in each host visited by the mobile code,
to protect this area is not easy.

\item\textit{Intelligent agents.\/} An intelligent agent is a software
object that have the ability of process information retrieved
autonomously. These agents does not require to be mobile. Intelligent
agents are an active field of study in \textit{artificial intelligence\/}
(AI) at present.
\end{itemize}

A given agent could be classified in more than one of the groups described
above. For example, a mobile agent could be programmed in a way that
allows it to decide the route that it will follow by evaluating the
information provided by the \textit{peer\/} (remote) hosts, making it both
a mobile and an intelligent agent simultaneously.

The threat we propose in our work allows an agent to be protected against
both malicious hosts and other agents. These hosts and agents could try to
make unauthorized modifications on either the code area or the data space
of the agent or even to remove the information provided by other hosts.
Our goal is to protect code and data areas against both counterfeit and
erasing. Our technique is based in the use of standard public-key
encryption algorithms ---also known as asymmetric ciphers--- instead of
symmetric ciphers. We will not propose nor recommend the use of a specific
cipher over others in our work.

We are not developing a protection algorithm for the execution thread. The
execution thread is changed by each host where the mobile code arrives. As
a consequence, it cannot be easily protected without using a logging
system. In our opinion, encryption techniques could not be used to provide
full protection of mobile code.

\section{Notational Conventions}

In this section, we will introduce the notation used in our work. This
notation will be applied to describe digital signature and encryption of
data in both the agent server and the peer hosts.

To describe the routes followed by the agents we will define the set
${\cal H}^{r} = \{{\rm H}_{1}^{r}, {\rm H}_{2}^{r}, \ldots,
{\rm H}_{n_{r}}^{r}\}$, where $r = 1,2,3,\ldots$ and
$n_{r}\in\mathbb{N}$; this set will denote the $n_{r}$ hosts followed by
the mobile agent released by a given agent server in its $r$-th route. In
this set, ${\rm H}_{i}^{r}$, where $i\in\{1, 2, \ldots, n_{r}\}$ and
$r\in\mathbb{N}$, determines the $i$-th host in the $r$-th route followed
by the agent sent by the server ${\rm S}$.

\textit{Digital signature\/}---$\!$ Digital signature of data can be used
to protect the information provided against counterfeiting and erasing by
allowing, at the same time, to be read and authenticated by other hosts.
It could be useful when agents are used in negotiation processes between
hosts. In this case, only the private/public key pairs related with the
remote hosts are used to protect the data stored in the agents allowing
the information to be read and authenticated by any host or agent without
a knowledge of the private-key used to digitally sign the message. We will
denote the digital signature of any given message ${\rm M}_{i, j}^{r}$
using the expression:
\begin{equation}
  {\rm S}_{i, j}^{r} \stackrel{\rm def}{=}
  f_{{\rm pri}_{{\rm H}_{i}^{r}}}
  \left[
    {\rm M}_{i, j}^{r}
  \right] \enspace ,
\end{equation}
where ${\rm M}_{i, j}^{r}$ and ${\rm S}_{i, j}^{r}$ are the plain-text
message and its signature respectively, ${\rm pri}_{{\rm H}_{i}^{r}}$
identifies the private-key associated with the host ${\rm H}_{i}^{r}$, and
$f$ stands for the digital signature algorithm applied to protect the
information provided by the peer hosts against unauthorized modifications.

The message ${\rm M}_{i, j}^{r}$ can be authenticated by using the
public-key associated with the $i$-th host in the agent route
${\cal H}^{r}$. That public-key may be available to any host. The security
of the host ${\rm H}_{i}^{r}$ is not compromised by storing a copy of that
public-key in a public-key server. As we will shown below, the use of
\textit{certification authorities\/} (CAs) to authenticate the public-keys
itself is highly recommended. We must apply
\begin{equation}
  {\rm M}_{i, j}^{r}
  = f_{{\rm pub}_{{\rm H}_{i}^{r}}}^{-1}
  \left[
    {\rm S}_{i, j}^{r}
  \right]
  = f_{{\rm pub}_{{\rm H}_{i}^{r}}}^{-1}
  \left[
    f_{{\rm pri}_{{\rm H}_{i}^{r}}}
    \left[
      {\rm M}_{i, j}^{r}
    \right]
  \right]
  \nonumber\enspace ,
\end{equation}
where ${\rm M}_{i, j}^{r}$ and ${\rm S}_{i, j}^{r}$ are the plain-text
message and the digital signature of ${\rm M}_{i, j}^{r}$ again,
${\rm pub}_{{\rm H}_{i}^{r}}$ stands for the public-key associated with
the host ${\rm H}_{i}^{r}$ and $f^{-1}$ identifies the digital signature
authentication algorithm.

\textit{Data encryption\/}---$\!$ A mobile agent based infrastructure
would require an additional security level. Suppose that a server drops an
agent that will retrieve information that must be covered to any host in
the network except the server that has released the agent itself. In this
case, data encryption should be used to hide the information stored in the
agent data area. Data encryption requires the use of an additional key
pair, the private/public key pair related with the agent server.
Encryption of data requires the use of the public-key provided by the
agent server to the remote hosts as a part of the mobile agent itself
(${\rm pub}_{\rm S}$), and the private-key associated with the remote host
that provides information to the agent. We will show below that the
private/public key pair related with the server that has released the
agent cannot be counterfeited without invalidating the agent itself
because it is a part of the code area. We will define the encryption
process as:
\begin{equation}
  {\rm C}_{i, j}^{r} \stackrel{\rm def}{=} f_{{\rm pub}_{\rm S}}
  \left[
    f_{{\rm pri}_{{\rm H}_{i}^{r}}}
    \left[
      {\rm M}_{i, j}^{r}
    \right]
  \right]
  \enspace .
  \label{eq:encryption}
\end{equation}
In this case, ${\rm M}_{i, j}^{r}$ and ${\rm C}_{i, j}^{r}$ are the
plain-text message and its cipher-text respectively. The symbols
${\rm pub}_{\rm S}$ and ${\rm pri}_{{\rm H}_{i}^{r}}$ in equation
(\ref{eq:encryption}) stand for the public-key related to the agent server
${\rm S}$ and the private-key associated with the $i$-th host in the
$r$-th route followed by an agent dropped by the server ${\rm S}$
respectively. In order to decrypt the cipher-text we must use the
private-key associated with the agent server (${\rm pri}_{\rm S}$), and
the public-key provided by the peer host that has encrypted the message
${\rm M}_{i, j}^{r}$:
\begin{eqnarray}
  {\rm M}_{i, j}^{r}
  &\!\!\!=\!\!\!& f_{{\rm pub}_{{\rm H}_{i}^{r}}}^{-1}
  \left[
    f_{{\rm pri}_{\rm S}}^{-1}
    \left[
      {\rm C}_{i, j}^{r}
    \right]
  \right]
  \label{eq:decryption} \\
  &\!\!\!=\!\!\!& f_{{\rm pub}_{{\rm H}_{i}^{r}}}^{-1}
  \left[
    f_{{\rm pri}_{\rm S}}^{-1}
    \left[
      f_{{\rm pub}_{\rm S}}
      \left[
        f_{{\rm pri}_{{\rm H}_{i}^{r}}}
        \left[
          {\rm M}_{i, j}^{r}
        \right]
      \right]
    \right]
  \right]
  \nonumber\enspace .
\end{eqnarray}
The elements that appear in equation (\ref{eq:decryption}) must be
interpreted in the same way as shown in other equations in this section.
In this case, ${\rm pri}_{\rm S}$ and ${\rm pub}_{\rm S}$ are,
respectively, the private and the public keys for the agent server;
${\rm pri}_{{\rm H}_{i}^{r}}$ and ${\rm pub}_{{\rm H}_{i}^{r}}$ are the
private and the public keys for the remote host ${\rm H}_{i}^{r}$. The
message that has been covered by using public-key cryptography is
${\rm M}_{i, j}^{r}$, that is the $j$-th message provided by the $i$-th
host in the route followed by the agent, and the cipher-text itself is
${\rm C}_{i, j}^{r}$.

\section{Code and Data Areas Protection}

Classical protection schemes do not allow a mobile agent to protect its
own code area against unauthorized modification easily. Suppose that a
server digitally signs the code of an agent before dropping it. This
server must provide copies of the public-key used to other hosts. This
public-key is required to authenticate the code area itself. This can be
done easily by providing a copy of this key to a key-server or by using a
CA. Obviously, the agent should be instructed to get this key showing at
least the address of both the server that has released it and the host
that stores a copy of the key. At this moment a malicious host, let us say
${\rm H}_{i}^{r,{\rm \ast}}$, could change the agent code area and sign it
by using its own private/public key pair. It is not difficult to prove
that this modification will not be discovered by the remote hosts if the
code is changed in such a way that the agent points to the new
public-key. The hostile host only needs to assure that the signed code
area provided by the agent server is recovered before the agent returns to
the server that has dropped it.

It is easy to see that code protection could not depend on classical
cryptography even if the keys used to authenticate this area are certified
and are provided by external trusted authorities. We need to develop a way
to link the data provided by the peer hosts with the code part of the
agent at the same time. This will allow us to protect data and code areas
simultaneously.

Other protection threats have been proposed in recent years. For example,
the use of both code and data areas mess up techniques as described in
\cite{hohl:mess}. In this reference, Hohl recommends the use of variable
names that does not means anything. He also proposed that the code should
not be modularized (\textit{i.e.\/}, it must be written without using
subroutines) and to choose a data representation that makes the program
difficult to understand. This author proposed the use of \textit{variable
recomposition\/} techniques, conversion of \textit{compile-time control
flow elements\/} into run-time data dependent jumps and the
\textit{insertion of dead code\/}, that is code that will not be executed
when the agent is running, into the agents. These techniques are based in
mixing up the contents of the variables and creating new variables that
contains a few bits of data of some of the original variables. These are
recovered by changing the way the code of the agent handles the access to
the variables. Another alternative could be to develop a secure
infrastructure for mobile agents \cite{yee:sanctuary}. The use of
\textit{encrypted functions\/}\footnote{By using mathematical functions
with homomorphic properties. An example is the exponential function where
the addition and mixed multiplication between $x, y \in \mathbb{C}$ could
be obtained without any explicit knowledge of $x$ providing ${\rm exp}
\left[ x \right]$ instead of $x$ in:
\begin{eqnarray}
  {\rm exp} \left[ x + y \right] &\!\!\!=\!\!\!&
    {\rm exp} \left[ x \right] \cdot {\rm exp} \left[ y \right] \nonumber\cr
  {\rm exp} \left[ x \cdot y \right] &\!\!\!=\!\!\!&
    \left( {\rm exp} \left[ x \right] \right) ^{y}
  \enspace ,
\end{eqnarray}
allowing us to represent the encrypted program as a polynomial. An
important requeriment is that these functions should not be easily
inverted. At present, there are not known one-way homomorphic functions.}
and \textit{execution environments\/} (EE) with a fully separated
interpreter\footnote{Where each agent have its own address space.} has
been proposed in \cite{sander:towards,tschudin:security}. At present, we
do not have a way of protecting the code against unauthorized modification
using encrypted functions because these techniques could not be easily
applied to real agent systems. We need protection schemes that do not rely
on hiding the algorithms used in code handling or on building trusted
environments for agent execution. Another problem is that the execution of
encrypted functions requires the development of one-way homomorphic
functions. These functions are unknown at present.

In order to protect both the message provided by the remote host, that we
have denoted by ${\rm M}_{i, j}^{r}$, and the mobile agent code area we
propose that each host must obtain the next field:
\begin{equation}
  {\rm M}_{{\rm CRC}_{i, j}}^{r} \stackrel{\rm def}{=}
  \underbrace{
    {\rm crc}
    \left[
      {\rm pub}_{\rm S}
    \right]
    + {\rm crc}
    \left[
      f_{{\rm pri}_{\rm S}}
      \left[
        {\rm M}_{\rm code}^{r}
      \right]
    \right]}_{\rm provided\,by\,the\,agent\,server\, S}
  + {\rm F}_{i, j}^{r}
  \enspace ,
  \label{eq:crc}
\end{equation}
where $f_{{\rm pri}_{\rm S}} \left[ {\rm M}_{\rm code}^{r} \right]$ stands
for the digital signature of the code area ${\rm M}_{\rm code}^{r}$. The
code area includes both the agent code and an identification number (ID)
for the agent. Therefore, this identification number is unique because it
is generated from the agent server identificator, which is unique in the
network (for example the IP-address in IPv4 or IPv6 format of the agent
server itself), and an agent number, which is unique for that server. In
our case, we will obtain ${\rm ID} = {\rm server}_{\rm ID} +
{\rm agent}_{\rm ID}$. Obviously, $f_{{\rm pri}_{\rm S}} \left[
{\rm M}_{\rm code}^{r} \right]$ could not be evaluated in the remote
hosts. To obtain this field, a knowledge of the private-key associated
with the server that has dropped the agent (${\rm pri}_{\rm S}$) is
required. This field must be provided as a part of the agent in a way that
cannot be falsified. To manage it, a field ${\rm F}_{i, j}^{r}$ must be
added by the host ${\rm H}_{i}^{r}$ to the field
${\rm M}_{{\rm CRC}_{i, j}}^{r}$ shown in equation (\ref{eq:crc}). This
field must change with each message provided in a way that it is unique
for that host/agent pair. The code part must be authenticated by comparing
its digital signature with the signature carried by the agent by each peer
host before accepting it. The \textit{cyclic redundancy check\/} (CRC) of
both the public-key associated with the agent server and the signature of
the agent code in (\ref{eq:crc}) can be obtained by each remote host by 
using a hash algorithm. This information must be matched with the CRCs of
both the public-key of the agent server and the digital signature provided
as a part of the data area.

In this section the term \textit{improved\/} (as appears in both
\textit{improved digital signature\/} and \textit{improved data
encryption\/}) stands for digital signature and encryption processes that
include information about the mobile agent code area. In fact, information
about the code part of the agent will be included by each peer host in the
field ${\rm M}_{{\rm CRC}_{i, j}}^{r}$, as shown in equation
(\ref{eq:crc}), allowing remote hosts to detect unauthorized modifications
of the code part of the agents. \textit{Partial data encryption\/} will
also provide a signed copy of the field ${\rm M}_{{\rm CRC}_{i, j}}^{r}$.
In this case, both the field ${\rm M}_{{\rm CRC}_{i, j}}^{r}$ and the
message ${\rm M}_{i, j}^{r}$ will be digitally signed but only the message
${\rm M}_{i, j}^{r}$ itself will be encrypted, allowing each host to
detect code tampering but protecting information against reading by
unauthorized hosts at the same time.

\textit{Improved digital signature\/}---$\!$ In our opinion it is possible
to protect both the agent code area and the information provided by the
host itself simultaneously. To manage it, the field
${\rm M}_{{\rm CRC}_{i, j}}^{r}$ must be used. This field must be stored
as a part of each message provided to the mobile agents before being
digitally signed:
\begin{equation}
  {\rm S}_{i, j}^{r} = f_{{\rm pri}_{{\rm H}_{i}^{r}}}
  \left[
    {\rm M}_{{\rm CRC}_{i, j}}^{r} + {\rm M}_{i, j}^{r}
  \right]
  \enspace ;
\end{equation}
this step assures data integrity while avoiding the possibility to
overwrite a new message provided by a remote host with an old one
provided to another mobile agent in the past. The information protected in
this way is authenticated by applying the public-key
${\rm pub}_{{\rm H}_{i}^{r}}$ of the host that has signed it:
\begin{eqnarray}
  {\rm M}_{{\rm CRC}_{i, j}}^{r} + {\rm M}_{i, j}^{r}
  &\!\!\!=\!\!\! & f_{{\rm pub}_{{\rm H}_{i}^{r}}}^{-1}
  \left[
    {\rm S}_{i, j}^{r}
  \right] \\
  &\!\!\!=\!\!\!& f_{{\rm pub}_{{\rm H}_{i}^{r}}}^{-1}
  \left[
    f_{{\rm pri}_{{\rm H}_{i}^{r}}}
    \left[
      {\rm M}_{{\rm CRC}_{i, j}}^{r} + {\rm M}_{i, j}^{r}
    \right]
  \right] \nonumber\enspace .
\end{eqnarray}

\textit{Improved data encryption\/}---$\!$ The information about the code
area of the mobile agent obtained by using (\ref{eq:crc}) will be
added to the message provided by the remote host. To encrypt a message for
the agent server we must apply the next algorithm to the message itself:
\begin{equation}
  {\rm C}_{i, j}^{r} = f_{{\rm pub}_{\rm S}}
  \left[
    f_{{\rm pri}_{{\rm H}_{i}^{r}}}
    \left[
      {\rm M}_{{\rm CRC}_{i, j}}^{r} + {\rm M}_{i, j}^{r}
    \right]
  \right] \enspace .
\end{equation}
In order to recover the full message, ${\rm M}_{{\rm CRC}_{i, j}}^{r} +
{\rm M}_{i, j}^{r}$, the agent server should apply its own private-key,
${\rm pri}_{\rm S}$, and the public-key provided by the host that has
encrypted the message, ${\rm pub}_{{\rm H}_{i}^{r}}$, obtaining:
\begin{equation}
  {\rm M}_{{\rm CRC}_{i, j}}^{r} + {\rm M}_{i, j}^{r} =
  f_{{\rm pub}_{{\rm H}_{i}^{r}}}^{-1}
  \left[
    f_{{\rm pri}_{\rm S}}^{-1}
    \left[
      {\rm C}_{i, j}^{r}
    \right]
  \right]
  \enspace .
\end{equation}
The main disadvantage of this method is that the code can be counterfeited
in such a way that only the agent server knows that it has been falsified.
If a part of ${\rm C}_{i, j}^{r}$ is not encrypted, but is digitally
signed, all the hosts in the route of the agent will have a way to
determine whether the code area or other data in the agent have been
falsified by any host.

\textit{Partial data encryption\/}---$\!$ A better answer to the problem
of data encryption is to provide information publically about the CRC of
both the public-key related with the agent server (${\rm pub}_{\rm S}$)
and the signature of the agent code, $f_{{\rm pri}_{\rm S}} \left[
{\rm M}_{\rm code}^{r} \right]$. It is possible to do it and, at the same
time, to hide the message in such a way that only the authorized host (the
agent server) could decrypt it. We propose to partially encrypt a
digitally signed message using:
\begin{equation}
  {\rm C}_{i, j}^{r} = f_{{\rm pri}_{{\rm H}_{i}^{r}}}
  \left[
    {\rm M}_{{\rm CRC}_{i, j}}^{r} + f_{{\rm pub}_{\rm S}}
    \left[
      {\rm M}_{i, j}^{r}
    \right]
  \right]
  \enspace ;
\end{equation}
assuring that the code area cannot be changed by a malicious host because
each one can authenticate the CRC of the digital signature, provided by
the agent server using (\ref{eq:crc}), and the public-key of that server
simultaneously. This information can be verified by any host but the
message ${\rm M}_{i, j}^{r}$ can only be decrypted using the private-key
of the agent server, ${\rm pri}_{\rm S}$. We propose to apply the next set
of equations to check the code and data areas of the agent and decrypt the
message:
\begin{eqnarray}
  {\rm M}_{{\rm CRC}_{i, j}}^{r} + f_{{\rm pub}_{\rm S}}
  \left[
    {\rm M}_{i, j}^{r}
  \right]
  &\!\!\!=\!\!\!& f_{{\rm pub}_{{\rm H}_{i}^{r}}}^{-1}
  \left[
    {\rm C}_{i, j}^{r}
  \right]
  \label{eq:certify}\\
  {\rm M}_{i, j}^{r}
  &\!\!\!=\!\!\!& f_{{\rm pri}_{\rm S}}^{-1}
  \left[
    f_{{\rm pub}_{\rm S}}
    \left[
      {\rm M}_{i, j}^{r}
    \right]
  \right] \enspace .
  \label{eq:addec}
\end{eqnarray}
The former equation (\ref{eq:certify}) can be used by any host because the
public-key related with the $i$-th host in the $r$-th route followed by an
agent sent by the server ${\rm S}$ (${\rm pub}_{{\rm H}_{i}^{r}}$) is
available publically to all the hosts that need it. This equation will be
required to check data and code integrity at the same time. Equation
(\ref{eq:addec}) is applied by the agent server using its own private-key,
${\rm pri}_{\rm S}$, to decrypt the information provided by the host ${\rm
H}_{i}^{r}$, after checking the cipher-text by applying
(\ref{eq:certify}).

\section{Public-key Propagation}

One of the main goals of our work is to provide a threat that allows
mobile agents to be protected against attacks like those described in
Section V-A. We propose to use public-key ciphers, also known as
asymmetric cryptosystems, instead of symmetric ciphers because the latter
allows a simplified key management in distributed environments.
Public-keys can be shared between hosts in a network without requiring
secure communication channels like these obtained using, for example, the
\textit{Transport Layer Security\/} (TLS) protocol. Detailed information
about the TLS protocol can be found in \cite{dierks:tls,lawrence:tls}.

Some important requeriments must be considered in the development of a
public-key propagation infrastructure for mobile agents:

\begin{itemize}
\item\textit{Certification authorities.\/} It is easy to see that
uncertified public-keys cannot be trusted. It is not a good practice to
send keys directly to the servers that need them. We need a network
infrastructure that allows the nodes to assure what host owns each
private/public key pair. For example middleman attack, the greatest known
vulnerability of public-key based ciphers, can be avoided by using a
trusted third party to verify and sign the keys transmitted over the
network.

\item\textit{Non-interactive protocol.\/} As pointed out it
\cite{sander:towards} a security model for mobile agents should conceive
protocols that require minimal interaction between the agent and the
server that has sent it. The server may want to go off-line, consequently,
the public-key should be provided by an independent host. Our threat
allows the public-key related with the agent server (${\rm pub}_{\rm S}$)
to be provided as a part of the agent.
\end{itemize}

Our threat to protect mobile agents offers some important advantages too.

\begin{itemize}
\item\textit{Secure communication channels are not required.\/} This is a
common advantage of public-key cryptography. As only public-keys are
transmitted over the network untrusted communication channels can be
established to share the keys. These keys cannot be used to falsify
information or decrypt data provided to the agents.

\item\textit{We do not need to know what host owns each public-key.\/}
Obviously, this fact is only true if different access privileges are not
assigned to each agent in function of the server that has released it. If
different access permissions are required CAs must be used to authenticate
the keys provided to the remote hosts and to assign the right access
privileges to each agent server.
\end{itemize}

Certified public-keys are required even if different access permissions
are not assigned to mobile agents in function of the server that has
dropped it. In fact, each peer host must identify other hosts in the
network in soon a way that it does not allows host impersonation
techniques. Trusted third parties are needed to avoid well known threats
like the middleman attack. We must consider that changing the public-keys
carried by mobile agents ---the public-keys associated with the agents
servers--- will invalidate data retrieved by the agents. If the
information stored in the mobile agent data area is removed from the agent
when this public-key is falsified other hosts will not have a way to
determine that the agent have been modified without authorization. But
this fact will be discovered by the agent server after the agent return.
These keys does not require to be authenticated using a CA. Changing
public-key for a given agent must be avoided once they have been provided
to the \textit{route servers\/} (RSs).

As we noted in Section III, both code and data areas can be protected
against counterfeiting and erasing by malicious hosts and agents. This can
be achieved by adding a field ${\rm M}_{{\rm CRC}_{i, j}}^{r}$ to each
message retrieved by a mobile agent as presented in (\ref{eq:crc}). This
field, ${\rm M}_{{\rm CRC}_{i, j}}^{r}$, will provide information about
both the code part of the agent and the public-key of its originating
server. This field will be stored in such a way that it does not allow
changing the code of the agent without invalidating it.

Each agent have its own ID to avoid the possibility of overwriting the
information provided by peer hosts with old data retrieved by other agents
sent by the same server. This field is stored as a part of the code area.
As a consequence, ${\rm crc} \left[ f_{{\rm pri}_{\rm S}} \left[
{\rm M}_{\rm code}^{r} \right] \right]$ changes when new agents are
dropped. Even obsolete information provided to the same agent in the past
cannot be used to cover new data. Each host must generate a field, we
called ${\rm F}_{i, j}^{r}$, unique for each message. All these fields
should be sent to the RSs:
\begin{eqnarray}
  {\rm F}_{i}^{r}
  & \!\!\!=\!\!\! & {\rm ID} + {\rm F}_{i, 1}^{r} + {\rm F}_{i, 2}^{r}
    + \ldots + {\rm F}_{i, m_{i}^{r}}^{r} \cr
  & \!\!\!=\!\!\! & {\rm ID} + \sum_{j=1}^{m_{i}^{r}} {\rm F}_{i, j}^{r}
  \enspace ,
  \label{eq:idmess}
\end{eqnarray}
where ${\rm ID}$ is the agent identification number mentioned above. The
field ${\rm F}_{i}^{r}$ must be sent to RSs digitally signed by applying:
\begin{equation}
  {\rm S}_{{\rm fields}_{i}}^{r} =
  f_{{\rm pri}_{{\rm H}_{i}^{r}}}
  \left[
    {\rm F}_{i}^{r}
  \right]
  \enspace ,
\end{equation}
using the private/public key pair for the host that provides that
information. Any host can check each message provided by ${\rm H}_{i}^{r}$
by using the ${\rm F}_{i, j}^{r}$ fields, where $i \in \{1, 2, \ldots,
n_{r}\}$ and $j = 1, 2, \ldots, m_{i}^{r}$, stored in ${\rm F}_{i}^{r}$.
Those fields must be authenticated by using:
\begin{equation}
  {\rm F}_{i}^{r} =
  f_{{\rm pub}_{{\rm H}_{i}^{r}}}^{-1}
  \left[
    {\rm S}_{{\rm fields}_{i}}^{r}
  \right] =
   f_{{\rm pub}_{{\rm H}_{i}^{r}}}^{-1}
  \left[
    f_{{\rm pri}_{{\rm H}_{i}^{r}}}
    \left[
      {\rm F}_{i}^{r}
    \right]
  \right]
  \enspace .
\end{equation}
To assure the integrity of the message ${\rm F}_{i}^{r}$ each RS should
send a random message to the host that wants to provide a new (or an
updated) message ${\rm F}_{i}^{r}$. This random message must be digitally
signed and returned to the RS that released it. The random message
signature must be checked before the RS accepts ${\rm F}_{i}^{r}$. In
other case, a malicious host can provide an obsolete message
${\rm F}_{i}^{r, {\rm \ast}}$ to the RSs overwriting the mobile agent data
area with old information corresponding to the false message
${\rm F}_{i}^{r, {\rm \ast}}$ simultaneously.

\section{Attacks against Mobile Agents Infrastructures}

In this section we classify the attacks that is possible to try against
the mobile agents and other hosts in the network. We show how our
protection threat allows us to protect the agents and, in some cases, even
remote hosts against these attacks.

\subsection{Attacks against Mobile Agents}

The main goal of our investigation is to protect the mobile agents against
both malicious hosts and other agents that can counterfeit the code part
and/or data areas of the agents. These hostile agents and hosts can try to
remove information carried by the agents too. As noted, these attacks
could arrive from both other agents and the hosts where the agents are
stored. In both cases, we will protect the agent using the same threat.

\subsubsection{Attacks against the Code Part}

As we shown in Section III the agent code area can be protected by adding
two CRCs to each message provided by the peer hosts. These CRCs provides
information that allows peer hosts to authenticate both the public-key
associated with the agent server and the digital signature of the code
area of the agent, that has been provided by the agent server itself.
These fields must be matched with each message stored in the data area by
the hosts followed in the agent trip, as appears in the set ${\cal
H}^{r}$. At last, each host authenticates the code area of the agent too
before running it using that digital signature previously checked. As both
the CRC related with the digital signature of the code area and the CRC
for the public-key of the agent server cannot be changed during agent
trip, but the latter is unique for a given agent, data area is protected
at the same time. As a consequence, the attacks described below can be
avoided.

\subsubsection{Attacks against the Data Area}

The attacks against the information carried by mobile agents can be
classified in three groups: (\textit{i\/}) attacks trying to erase data
carried by the agents (also known as a mobile agent ``brainwashing'' in
bibliography), (\textit{ii\/}) attacks trying to falsify data provided by
other hosts and (\textit{iii\/}) attacks trying to uncover non-public
information carried by the agents.

\textit{Removing information\/}---$\!$ To avoid a mobile agent
``brainwash'' each host must provide information about the number of
messages stored in a particular agent to the RSs as we described in
Section IV. It is easy to see that all the information needed to protect
the data area cannot be stored in the agent. Suppose for example that a
mobile agent carries a set of signed messages
\begin{eqnarray}
{\cal D}^{r} = \{
  {\rm S}_{1, 1}^{r}, {\rm S}_{1, 2}^{r}, \ldots, {\rm S}_{1, m_{1}^{r}}^{r},
  {\rm S}_{2, 1}^{r}, {\rm S}_{2, 2}^{r}, \ldots, {\rm S}_{2, m_{2}^{r}}^{r},
  \ldots, \cr
  {\rm S}_{i, 1}^{r}, {\rm S}_{i, 2}^{r}, \ldots, {\rm S}_{i, j}^{r}, \ldots,
  {\rm S}_{i, m_{i}^{r}}, \ldots, \cr
  {\rm S}_{n_{r}, 1}^{r}, {\rm S}_{n_{r}, 2}^{r},
  \ldots, {\rm S}_{n_{r}, m_{n_{r}}^{r}}^{r} \}
  \enspace ,
  \label{eq:dasgn}
\end{eqnarray}
in its data area. The set ${\cal D}^{r}$ in (\ref{eq:dasgn}) stands for
the signed data area of the $r$-th agent released by a server. All the
messages are signed in a way that only the hosts that has provided these
messages can change its contents. If the RSs do not provides information
about the number of messages given by each host to the agent server or,
more generally to other hosts that requests it, any hostile node (any
malicious host in the network) can remove a message, let us say
${\rm S}_{i, j}^{r}$ where $i \in \{1, 2, \ldots, n_{r}\}$ identifies the
host that provides the message removed and $j \in \{1, 2, \ldots,
m_{i}^{r}\}$ stands for the $j$-th message provided by that host. In this
case, the set of messages carried by the agent, ${\cal D}^{r}$, is changed
to
\begin{eqnarray}
{\cal D}^{r, {\rm \ast}} = \{
  {\rm S}_{1, 1}^{r}, {\rm S}_{1, 2}^{r}, \ldots, {\rm S}_{1, m_{1}^{r}}^{r},
  {\rm S}_{2, 1}^{r}, {\rm S}_{2, 2}^{r}, \ldots, {\rm S}_{2, m_{2}^{r}}^{r},
  \ldots, \nonumber\cr
  {\rm S}_{i, 1}^{r}, {\rm S}_{i, 2}^{r},
  \stackrel{\wedge ^{{\rm S}_{i, j}^{r}}}{\ldots},
  {\rm S}_{i, m_{i}^{r}}, \ldots, \cr
  {\rm S}_{n_{r}, 1}^{r}, {\rm S}_{n_{r}, 2}^{r}, \ldots,
  {\rm S}_{n_{r}, m_{n_{r}}^{r}}^{r} \}
  \enspace ,
\end{eqnarray}
without invalidating the data area. In this case, the set
${\cal D}^{r, {\rm \ast}}$ is the falsified data area of the agent. The
same problem happens when encryption is used if the number of messages
carried by a mobile agent is not provided to the RSs in any way. In this
case the encrypted data area of the agent:
\begin{eqnarray}
{\cal D}^{r} = \{
  {\rm C}_{1, 1}^{r}, {\rm C}_{1, 2}^{r}, \ldots, {\rm C}_{1, m_{1}^{r}}^{r},
  {\rm C}_{2, 1}^{r}, {\rm C}_{2, 2}^{r}, \ldots, {\rm C}_{2, m_{2}^{r}}^{r},
  \ldots, \cr
  {\rm C}_{i, 1}^{r}, {\rm C}_{i, 2}^{r}, \ldots, {\rm C}_{i, j}^{r}, \ldots,
  {\rm C}_{i, m_{i}^{r}}, \ldots, \cr
  {\rm C}_{n_{r}, 1}^{r}, {\rm C}_{n_{r}, 2}^{r},
  \ldots, {\rm C}_{n_{r}, m_{n_{r}}^{r}}^{r} \}
  \enspace ,
  \label{eq:daencr}
\end{eqnarray}
can be modified by removing one of the cipher-texts provided by the remote
hosts. For example, the cipher-text that corresponds to the $j$-th message
provided by the $i$-th host in the agent route can be erased by changing
the agent data area to:
\begin{eqnarray}
{\cal D}^{r, {\rm \ast}} \!=\! \{
  {\rm C}_{1, 1}^{r}, {\rm C}_{1, 2}^{r}, \ldots, {\rm C}_{1, m_{1}^{r}}^{r},
  {\rm C}_{2, 1}^{r}, {\rm C}_{2, 2}^{r}, \ldots, {\rm C}_{2, m_{2}^{r}}^{r},
  \ldots, \nonumber\cr
  {\rm C}_{i, 1}^{r}, {\rm C}_{i, 2}^{r},
  \stackrel{\wedge ^{{\rm C}_{i, j}^{r}}}{\ldots},
  {\rm C}_{i, m_{i}^{r}}, \ldots, \cr
  {\rm C}_{n_{r}, 1}^{r}, {\rm C}_{n_{r}, 2}^{r}, \ldots,
  {\rm C}_{n_{r}, m_{n_{r}}^{r}}^{r} \}
  \enspace .
\end{eqnarray}
Even if information about the number of messages carried by the agent is
provided in a way that an hostile peer host cannot change it, data area
can be altered by overwriting it with a \textit{bit-copy\/} of an old data
area.

To avoid attacks based in the techniques described above we propose to
send information about the number of messages provided to the agent to the
RSs shown in the code area of the mobile agent itself. These hosts cannot
be changed without invalidating the agent itself. A copy of the fields
${\rm F}_{i, j}^{r}$, as shown in equation (\ref{eq:crc}), is all the
information needed to manage it; in fact, these fields are required to
authenticate data provided by each host visited by the mobile agent as
shown above.

\textit{Counterfeit of data\/}---$\!$ Even if data have been digitally
signed or encrypted information provided can be falsified. As noted above,
data area can be protected against ``brainwashing'' by storing information
about the amount of messages provided by each host visited by the agent in
separated hosts. We need a way to protect the information provided by peer
hosts to the agent against being overwritten with old signed data provided
by those hosts in the past. In Section III we propose to link data
provided as a part of the agent, the fields in
${\rm M}_{{\rm CRC}_{i, j}}^{r}$ as shown in (\ref{eq:crc}), with each
messages retrieved by the agent. If this field is not included any hostile
host can change the data area of the agent as appears in (\ref{eq:dasgn}),
overwriting a valid signed message ${\rm M}_{i, j}^{r}$ with an old
message signed by the same host using the same private/public key pair,
let us say ${\rm M}_{i, j}^{r, {\rm \ast}}$:
\begin{eqnarray}
  {\cal D}^{r, {\rm \ast}} = \{
  {\rm S}_{1, 1}^{r}, {\rm S}_{1, 2}^{r}, \ldots, {\rm S}_{1, m_{1}^{r}}^{r},
  {\rm S}_{2, 1}^{r}, {\rm S}_{2, 2}^{r}, \ldots, {\rm S}_{2, m_{2}^{r}}^{r},
  \ldots, \nonumber\cr
  {\rm S}_{i, 1}^{r}, {\rm S}_{i, 2}^{r}, \ldots,
  {\rm S}_{i, j}^{r,{\rm \ast}}, \ldots, {\rm S}_{i, m_{i}^{r}}, \ldots, \cr
  {\rm S}_{n_{r}, 1}^{r}, {\rm S}_{n_{r}, 2}^{r},
  \ldots, {\rm S}_{n_{r}, m_{n_{r}}^{r}}^{r} \}
  \enspace .
\end{eqnarray}
The same problem happens with encrypted messages if the field ${\rm
M}_{{\rm CRC}_{i, j}}^{r}$, obtained by applying (\ref{eq:crc}), is not
provided as a part of the messages. In this case, the encrypted data area
of the mobile agent in (\ref{eq:daencr}) can be counterfeited by changing
one of the cipher-texts provided by the remote hosts, for example the
message ${\rm C}_{i, j}^{r}$ can be changed to
${\rm C}_{i, j}^{r, {\rm \ast}}$:
\begin{eqnarray}
{\cal D}^{r, {\rm \ast}} \!=\! \{
  {\rm C}_{1, 1}^{r}, {\rm C}_{1, 2}^{r}, \ldots, {\rm C}_{1, m_{1}^{r}}^{r},
  {\rm C}_{2, 1}^{r}, {\rm C}_{2, 2}^{r}, \ldots, {\rm C}_{2, m_{2}^{r}}^{r},
  \ldots, \nonumber\cr
  {\rm C}_{i, 1}^{r}, {\rm C}_{i, 2}^{r}, \ldots,
  {\rm C}_{i, j}^{r,{\rm \ast}}, \ldots, {\rm C}_{i, m_{i}^{r}}, \ldots, \cr
  {\rm C}_{n_{r}, 1}^{r}, {\rm C}_{n_{r}, 2}^{r},
  \ldots, {\rm C}_{n_{r}, m_{n_{r}}^{r}}^{r} \}
  \enspace .
\end{eqnarray}
To avoid information provided by peer hosts to be overwritten by using a
bit-copy with old signed data obtained in the same trip we propose to add
another field to ${\rm M}_{{\rm CRC}_{i, j}}^{r}$. This field is
introduced in Section III. Each host can provide a field in each message.
We denoted this field as ${\rm F}_{i, j}^{r}$. This field is changed when
the remote host wants to sign or encrypt a new message for the agent. This
field will be provided to the RSs and must be matched against the copies
stored in ${\rm F}_{i, j}^{r}$ by each host that wants to check data and
code integrity. Each host visited by the agent provides a set of fields
${\rm F}_{i, j}^{r}$, where $j = 1, 2, \ldots, m_{i}^{r}$ to the RSs as
presented in (\ref{eq:idmess}).

\textit{Cryptanalysis\/}---$\!$ In our work we propose to protect data
provided by using standard cryptographic techniques that can be attacked
by using cryptanalysis\footnote{Two powerful techniques to attack ciphers,
known as \textit{differential\/} and \textit{linear\/} cryptanalysis, were
elaborated and are being currently used. The former was developed by Adi
Shamir and Eli Biham of Technion Israel Institute of Technology. The
latter was introduced by Mitsuru Matsui of Mitsubishi Electric
Corporation.}. We are not making assumptions about the encryption
algorithms used to cover data nor the keys length that may vary in
function of the security requirements. As noted in
\cite{zimmermann:cryptography}, the fact that public-key based ciphers
allows predictable patterns to survive the encryption process making this
technology vulnerable to cryptanalysis is well known to cryptanalysts; as
a consequence, standard compression techniques should be applied before
encryption to increase data security.

\textit{Middleman attack\/}---$\!$ The man-in-the-middle attack is
probably the greatest known vulnerability of asymmetric cryptosystems.
Mobile agent based infrastructures can be attacked by using a middleman
attack variant. In this case, a malicious host will intercept both the
public-key send to the RSs by the remote host and the agent itself. This
host will generate a private/public key pair to falsify data provided by
that host and provide a copy of the false public-key to the RSs. To avoid
the attacks based on this threat the use of CAs to verify and sign the
keys used by the hosts to protect its own data is recommended.

\subsubsection{Attacks against the Agents itself}

The mobile agents can be attacked in a way that do not require to modify
either the code area or the data part of the agents. The main goal of
these attacks can be to damage the agent infrastructure itself by
destroying the agents or releasing new agents instead of the original one.

\textit{Removing agents\/}---$\!$ Any malicious host can remove the agents
when arrive to it. There are no-way to avoid this attack against the
agents but or protection threat allows other host to try to discover what
host has killed the agent by requesting information about the route
followed by the agent.

\textit{Releasing new agents\/}---$\!$ A hostile host can remove all the
information stored in the mobile agent and change the code area. Modifying
the code area requires gathering a fake private/public key pair for the
agent server but now it is possible because all the
${\rm M}_{{\rm CRC}_{i, j}}^{r}$ fields have been removed from the data
area of the agent. The new private/public key pair can be used to sign a
modified code area of the agent. This fact can be discovered, at least, by
the server that has released the agent when it comes back. If other hosts
have a copy of the public-key of the agent server, either obtained by
other channels or sent in the past, these hosts can discover the
unauthorized modification of the agent too.

\subsection{Attacks against Peer Hosts}

Our goal is to protect mobile agents against attacks from both peer hosts
and other agents. We are not trying to develop a threat to protect hosts
against malicious agents. Attacks against remote hosts could be initiated
from both agents and hosts.

The code area protection allows an agent to be protected against malicious
changes that could affect how it works. At the same time, the code
protection allows a host to be protected against \textit{Denial of
Service\/} (DoS) attacks by agent cloning in the sense that the number of
clones could be easily verified by using the agent identification number
described above. This allows a host to protect itself by controlling the
resources provided to the agents in a \textit{per-agent\/} basis. The
agent identification number allows a host to identify the number of clones
of a given agent.

The code protection threat proposed in our work do not permits a hostile
host to change the code part of an agent provided by an agent server
without invalidating it but, obviously, this host could release its own
malicious agents.

\section{Conclusions}

Mobile agents are an extremely vulnerable piece of software because they
are executed in untrusted environments. Both code and data areas must be
protected against malicious hosts and agents. The former requires
techniques that does not allow a malicious host to hide the identity of
the real agent owner. The latter requires information provided by remote
hosts to be protected against counterfeit and erasing. The main advantages
of the algorithm proposed in this paper are that:

\begin{itemize}
\item\textit{Secure communication channels are not required\/} allowing a
mobile agent to be transmitted over untrusted channels and even stored in
malicious hosts where the agent will be shown as plain-text even if
trusted communication channels between hosts are established.

\item\textit{Both code and data areas are protected\/} against counterfeit
and erasing; consequently, mobile agents are a more secure and robust
platform.

\item\textit{Each host could change its own information\/} when required.
This allows a host to update information provided permitting the
development of more sophisticated agent-based applications, where
negotiation between agents and hosts is required.
\end{itemize}

We hope that our protection scheme allows mobile agent based
infrastructures to be protected against other attacks based on threats
not covered in this article or even unknown at present. If this can be
achieved, our threat could be a good design principle for mobile agent
based information networks.

\section*{Acknowledgments}

The authors would like to thank Dr. Agust\'{\i}n Nieto, Dr. Jos\'e Manuel
Noriega and Dr. M.~\'A.~R. Osorio for reviewing the draft of the article,
recommend us the use of the notation proposed by the \textit{American
Mathematical Society\/} (AMS) in this work and provide us a place to work.
Without their many helpful comments this work would not be possible.

\nocite{*}
\bibliographystyle{IEEE}

\begin{thebibliography}{10}

\bibitem{ritchie:security}
Dennis~M. Ritchie,
\newblock {\em On the Security of UNIX},
\newblock Unix Programmer's Manual. A. G. Hume and M. D. McIlroy, AT{\&}T Bell
  Laboratories, Murray Hill, N. J., June 1977.

\bibitem{hohl:mess}
Fritz Hohl,
\newblock ``An approach to solve the problem of malicious hosts in mobile
  agents systems,''
\newblock {\em Institute of Parallel and Distributed High-Performance Systems
  (IPVR), University of Stuttgart, Germany}, vol. 1997, no. 03, pp. 1--13,
  March 1997.

\bibitem{sander:hosts}
Tomas Sander and Christian~F. Tschudin,
\newblock ``Protecting mobile agents against malicious hosts,''
\newblock {\em Lecture Notes in Computer Science (LNCS), Springer-Verlag Inc.,
  New York, NY, USA}, vol. 1419, June 1998.

\bibitem{sobrado:otp}
Igor Sobrado,
\newblock {\em A One-Time Pad based Cipher for Data Protection in Distributed
  Environments},
\newblock [Online], Computing Research Repository (CoRR), arXiv:cs.CR/0005026,
  Available: http://xxx.lanl.gov/abs/cs.CR/0005026, May 2000.

\bibitem{yee:sanctuary}
Bennet~S. Yee,
\newblock ``A sanctuary for mobile agents,''
\newblock {\em Secure Internet Programming, Lecture Notes in Computer Science
  (LNCS), Springer-Verlag Inc.}, vol. 1603, pp. 261--274, 1999.

\bibitem{sander:towards}
Tomas Sander and Christian~F. Tschudin,
\newblock ``Towards mobile cryptography,''
\newblock {\em International Computer Science Institute (ICSI) Technical
  Report}, vol. 97, no. 049, pp. 1--14, November 1997.

\bibitem{tschudin:security}
Christian~F. Tschudin,
\newblock ``Intelligent Information Agents --- Agent based information
  discovery and management on the Internet,''
\newblock in {\em Lecture Notes in Computer Science (LNCS), Springer-Verlag
  Inc., New York, NY, USA}, M.~Klusch, Ed., July 1999, pp. 431--445.

\bibitem{dierks:tls}
T.~Dierks and C.~Allen,
\newblock ``The TLS protocol: Version 1.0,''
\newblock {\em Request for Comments}, , no. 2246, pp. 1--80, January 1999.

\bibitem{lawrence:tls}
Lawrence~C. Paulson,
\newblock ``Inductive analysis of the Internet protocol TLS,''
\newblock {\em ACM Transactions on Information and System Security}, vol. 2,
  no. 3, pp. 332--351, August 1999.

\bibitem{zimmermann:cryptography}
Philip~R. Zimmermann,
\newblock ``Cryptography for the Internet,''
\newblock {\em Scientific American}, vol. 279, no. 4, pp. 82--87, October 1998.

\end{thebibliography}

\end{document}